\documentclass[sigconf,final]{acmart}

\AtBeginDocument{%
  \providecommand\BibTeX{{%
    \normalfont B\kern-0.5em{\scshape i\kern-0.25em b}\kern-0.8em\TeX}}}

\renewcommand\footnotetextcopyrightpermission[1]{} 
\acmISBN{978-1-4503-XXXX-X/18/06}
\setcopyright{none}              %
\acmConference[]{}{}{}           %
\acmYear{}                       %
\acmDOI{}                        %
\acmISBN{}                       %

\acmSubmissionID{2997}

\begin{document}

\title{Towards Interactive Multimodal Representation of ML Functions for Human Understanding of ML}


\author{Bokang Wang}
\affiliation{
    \institution{Carnegie Mellon University}
    \city{Pittsburgh}
    \state{PA}
    \country{USA}
}

\author{Yingxuan Liao}
\affiliation{
    \institution{Carnegie Mellon University}
    \city{Pittsburgh}
    \state{PA}
    \country{USA}
}

\author{Leah Lee}
\affiliation{
    \institution{Carnegie Mellon University}
    \city{Pittsburgh}
    \state{PA}
    \country{USA}
}

\author{Jack Wesson}
\affiliation{
    \institution{Carnegie Mellon University}
    \city{Pittsburgh}
    \state{PA}
    \country{USA}
}

\author{Anlan Yang}
\affiliation{
    \institution{Carnegie Mellon University}
    \city{Pittsburgh}
    \state{PA}
    \country{USA}
}

\author{Ruizi Wang}
\affiliation{
    \institution{Carnegie Mellon University}
    \city{Pittsburgh}
    \state{PA}
    \country{USA}
}

\author{Yigang Wen}
\affiliation{
    \institution{Carnegie Mellon University}
    \city{Pittsburgh}
    \state{PA}
    \country{USA}
}





\begin{abstract}
Attitudes about artificial intelligence and machine learning are recent victims of endemic misunderstanding; given our increasing reliance on these technologies, the need for widespread understanding and confidence in their use is paramount\cite{Umang20}. To this end, our work seeks to increase understanding in these typically inaccessible topics through interactive visualizations\cite{Claire19}, thereby garnering curiosity in the hopes of kickstarting a cycle of understanding leading to further pursuit of knowledge. We hope this will cyclically shift global attitudes away from the intimidation of the unknown currently plaguing ML. This work explores best practices for supporting curiosity in new technologies, to inspire attitudinal paradigm-shifts. Over three, distinct visualizations of machine learning data, we created prototypes with carefully selected, highly-transparent datasets, to examine the success factors of engagement required for more informed attitudes on ML less dictated by the fear of the unknown. By employing interactive visualizations, we can captivate the interest of teenagers and individuals from diverse fields, encouraging them to explore the fascinating world of machine learning.
\end{abstract}



\begin{teaserfigure}
  \includegraphics[width=\textwidth,height=4cm]{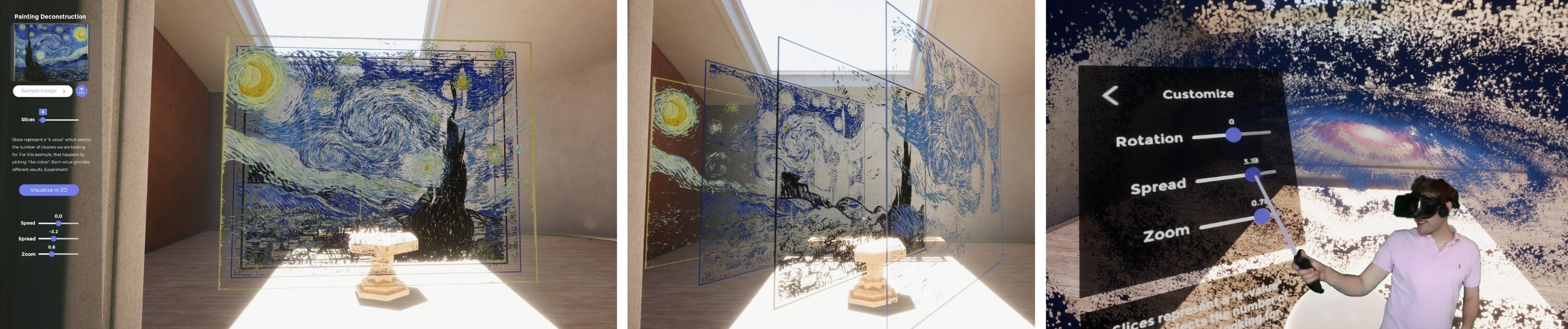}
  \caption{K-Means clustering deconstructs user paintings to show clustering in 3D and VR}
  \label{fig:teaser}
  \Description{K-Means clustering deconstructs user paintings to show clustering in 3D and VR}
\end{teaserfigure}


\maketitle

\section{Introduction and Motivation}
 This is a product development project to discover how 3D data visualization and interactions can be used to represent various features of machine learning algorithms in an engaging environment. Our research was motivated by the “black box” perceptions of ML in average audiences, and those who find it intimidating from lack of knowledge are our target audience.
     
     We developed three visualizations: 1. analyzing RGB datasets using proximity clustering algorithms on image’s pixel colors, 2. exploring novelty in interaction using haptic feedback to deconstruct songs, and 3. shaping reinforced learning experiences using relatable metaphors in narrative storytelling. Through our development, we discovered that engaging experiences foster a greater inclination to explore the “unknown”, building a cyclical relationship between the pursuit of learning and resultant engagement.

\section{Isochromatic Deconstruction}

\subsection{Introduction}

In this prototype, we embarked on an endeavor to delve into the intricate depths of visual data by employing the K-means clustering algorithm to dissect RGB datasets. Our primary aim was to deconstruct paintings into isochromatic layers, unraveling the underlying color complexities that constitute these artistic masterpieces. Leveraging the RGB values as an XYZ dataset, we orchestrated the clustering of images into layers meticulously organized based on color proximity. This innovative approach transcends the conventional two-dimensional realm, propelling the visualization process into the captivating realms of a three-dimensional space. By seamlessly integrating Unity 3D as our development platform, we crafted an immersive environment where the boundaries between reality and digital art blur. The culmination of our efforts materialized in the form of a virtual space, brought to life through the Oculus Quest 2 Head Mounted Display and trackers. This amalgamation of technology allowed us to immerse ourselves and our audience in a multisensory journey, where every pixel of the painting pulsated with vibrancy and depth. Through this transformative experience, we not only visualized static images but also breathed life into them, fostering a profound connection between art and observer.

\subsection{Result}
Numerous elements contributed to the success of our approach. Firstly, the visualization of direct and unequivocal data played a pivotal role. The inherent clarity of a 2D image transformed into a lucid pixel dataset, facilitating a seamless understanding of the underlying information and making a deep impression on the viewer's mind. This clarity not only enhanced comprehension but also heightened the overall impact of the experience, ensuring that every detail of the artwork was vividly appreciated. Furthermore, the incorporation of customization features proved instrumental in fostering a sense of personal investment among participants. By empowering users to select their preferred images for analysis, we created a more intimate and relatable interaction, wherein individuals felt a deeper connection to the visualized data. This personalization not only enriched the experience but also instilled a sense of ownership, thereby enhancing engagement levels. Moreover, the integration of virtual reality (VR) technology served as a catalyst for immersion, propelling users into an alternate dimension where the boundaries between reality and digital art dissolved. The immersive nature of the VR variant provided users with an array of sensory stimuli, captivating their attention and imbuing the experience with a heightened sense of realism. Consequently, participants were not merely passive observers but active participants in a visually captivating journey, thereby elevating the overall impact and memorability of the experience. 

\section{Haptic Music}
\begin{figure}[ht]
  \centering
  \includegraphics[width=\linewidth]{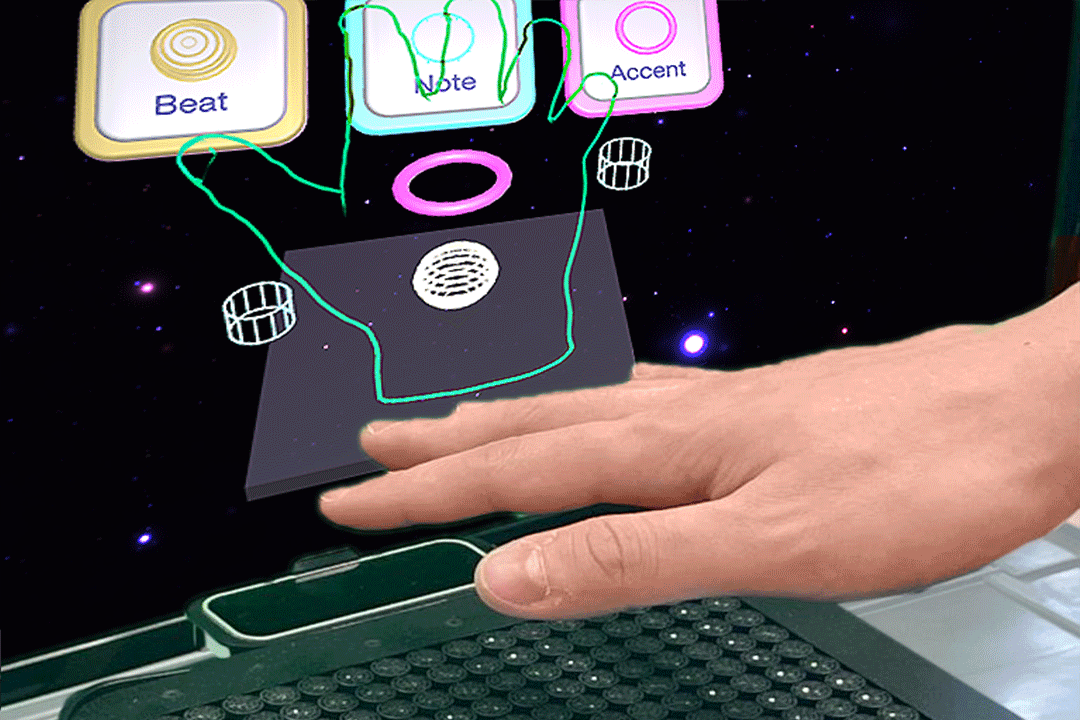}
  \caption{Musical subcomponent analysis used the Fast Fourier transform algorithm to experience music haptically}
  \Description{A human hand over the haptics device}
\end{figure}
\subsection{Introduction}
Our approach involved intricately mapping the beats, notes, and accents of a song onto distinct sensations dispersed across the hand, thereby forging an unprecedented connection between auditory stimuli and tactile feedback. We want not only visualizing the music but also making the music touchable. This dynamic fusion of sensory modalities not only enabled us to craft mesmerizing visual metaphors on screen but also synchronized sound and haptics to heighten the visceral experience of 'feeling' music. 

Firstly, our methodology entails employing the fast Fourier transform algorithm to meticulously dissect the beats, notes, and intensity of the music. This algorithm enables us to extract intricate details from the audio, facilitating a comprehensive understanding of its rhythmic and tonal nuances.

Once we have obtained the essential beats, notes, and intensity data, our next step involves translating this information into a visually captivating representation on the screen. Through intuitive visualization techniques, users will be able to perceive the underlying musical structure and dynamics with clarity and precision.

Furthermore, we aim to enhance the user experience by integrating the STRATOS Ultrahaptics\cite{Ultraleap22} technology. Developed by Leap Motion, STRATOS Ultrahaptics harnesses the power of sonic vibration to generate tactile sensations without physical contact. By synchronizing with Leap Motion's hand-tracking capabilities, this innovative device can imbue users with haptic feedback, enabling them to feel the music in a profoundly immersive manner. 
\subsection{Steps}
In our approach to handling the notes within the music, we adopt a systematic mapping strategy to ensure efficient and intuitive interaction. Given that there are 12 distinct types of notes in total, we allocate each note to specific fingers, enabling users to navigate through the musical landscape with ease and precision.

Thumb: This versatile digit is assigned a range of notes, including C, C\#, A, and A\#. By utilizing the thumb, users can effortlessly access these fundamental notes, laying the foundation for their musical exploration.
Index Finger: Positioned next in line, the index finger is entrusted with the notes D, D\#, and B. 
Middle Finger: Singularly dedicated to the note E.
Ring Finger: With the responsibility of representing the notes F and F\#, the ring finger assumes a vital role in our mapping strategy. 
Little Finger: Completing our comprehensive mapping scheme, the little finger is designated to handle the notes G and G\#. 

we implement a transparent hand visualization on the screen, allowing users to observe in real-time how beats, notes, and accents hit and align with their virtual hand. Moreover, our commitment to enhancing user engagement extends beyond mere visual representation. We strive to synchronize the haptic feedback provided by the STRATOS Ultrahaptics device with the corresponding visual elements displayed on the screen. This integration ensures that users not only see but also feel the rhythmic pulse, notes, and accents as they interact with the virtual hand.

\begin{figure}[ht]
  \centering
  \includegraphics[width=\linewidth]{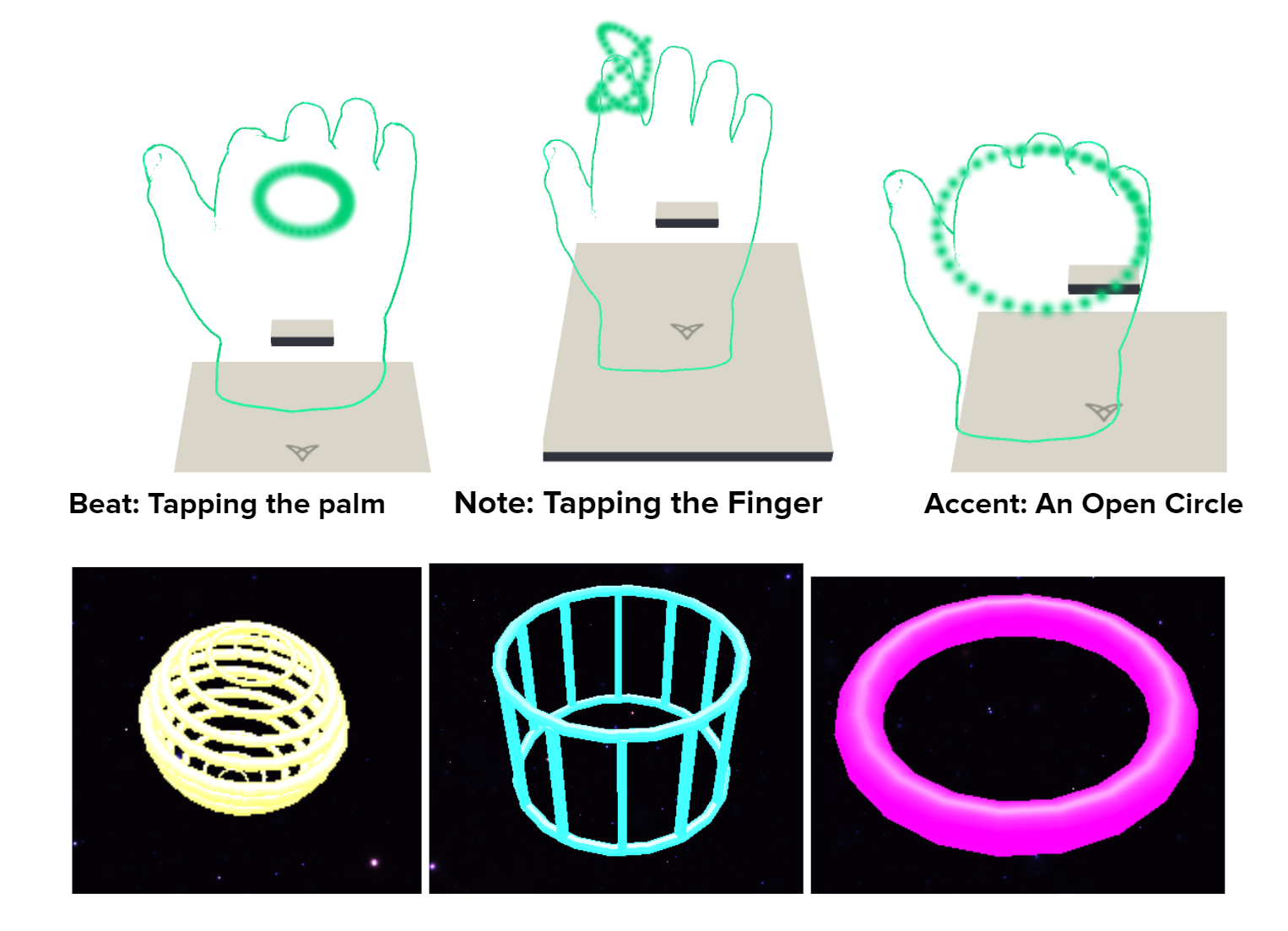}
  \caption{Comparison of the haptic feeling and the model of beat, note, and accent}
  \Description{Hand models and signs}
\end{figure}
Additionally, we meticulously crafted three tutorial levels to immerse the user in experiencing the rhythm, notes, and accents individually. We deliberately selected straightforward beats and melodies, ensuring that users can readily grasp the essence of our prototype. These tutorial levels serve as a comprehensive learning platform, allowing users to familiarize themselves with pressing buttons effectively.

\subsection{Result}

By harnessing the capabilities of STRATOS Ultrahaptics, users were transported into a realm where they could not only hear but also physically sense the intricate nuances of their favorite songs. 

The sheer novelty of this interaction paradigm captured the imagination of users from the outset, eliciting a level of interest that transcended conventional boundaries. However, it was not merely novelty alone that sustained engagement; rather, it was the innate returnability of the experience that fostered a lasting connection. Subsequent re-engagements with the platform served to deepen users' understanding and appreciation of the underlying mechanics, paving the way for a gradual yet profound evolution in their interaction with music. This iterative process of exploration and discovery underscored the transformative potential of our approach, as users embarked on a journey of sonic exploration that continually unfolded new layers of insight and understanding. Through this pioneering fusion of art and science, we not only expanded the horizons of musical expression but also redefined the very essence of what it means to 'feel' music. 


\section{Reinforcement Learning through Narrative}

Reinforcement learning stands as a cornerstone within the realm of machine learning methodologies, leveraging a paradigm rooted in rewarding desirable actions while penalizing undesirable ones. When visualizing the intricate process of reinforcement learning, we hone in on its fundamental components: action, state, reward, and evolution. Among the array of algorithms available, we've elected to employ Q-learning as our algorithmic backbone. Renowned for its simplicity and efficacy, Q-learning represents a model-free reinforcement learning approach geared towards comprehending the value associated with actions within specific states.

At its core, Q-learning operates on the premise of iteratively updating action-value functions, encapsulating the essence of decision-making in an uncertain environment. This dynamic process involves navigating through a space of states and actions, constantly refining strategies based on received rewards. The allure of Q-learning lies in its ability to adapt and evolve over time, learning optimal policies through trial and error without requiring a pre-existing model of the environment.

\begin{figure}[ht]
  \centering
  \includegraphics[width=\linewidth]{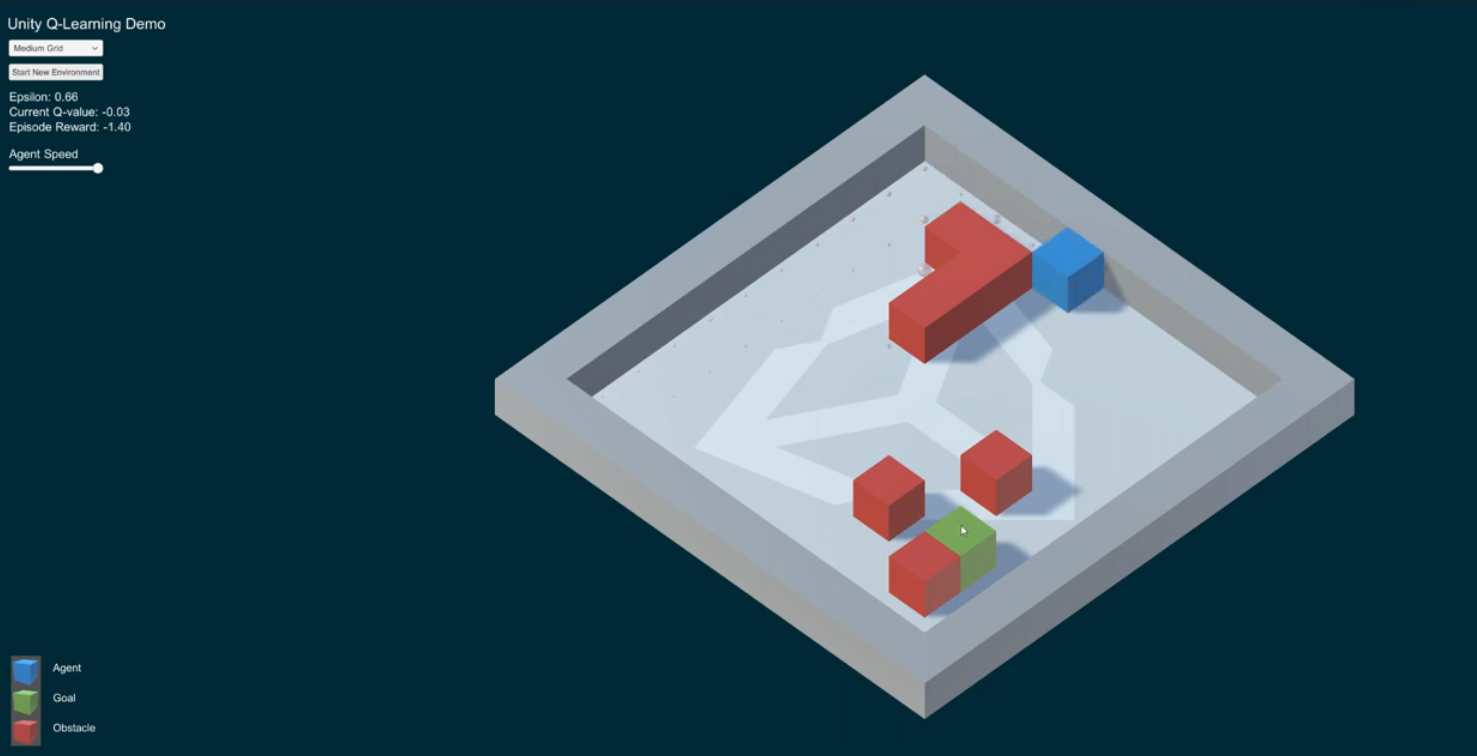}
  \caption{ The blue box: AI. The green box: goal. The red box: obstacle. The user can set up the environment and see how AI is trained to find the shortest path to reach the goal.}
  \Description{several read, blue cubes}
\end{figure}

Our rationale behind selecting Q-learning transcends its simplicity; crucially, it boasts the capability to operate in real-time. This attribute holds paramount significance, particularly in scenarios where immediacy and responsiveness are paramount. By harnessing Q-learning, we afford players the opportunity to witness the evolution of AI firsthand, as it swiftly adapts and refines its decision-making prowess in real-time. This seamless integration of algorithmic sophistication with real-world applicability underscores the profound impact and potential of reinforcement learning methodologies within diverse domains.

\begin{figure}[ht]
  \centering
  \includegraphics[width=\linewidth]{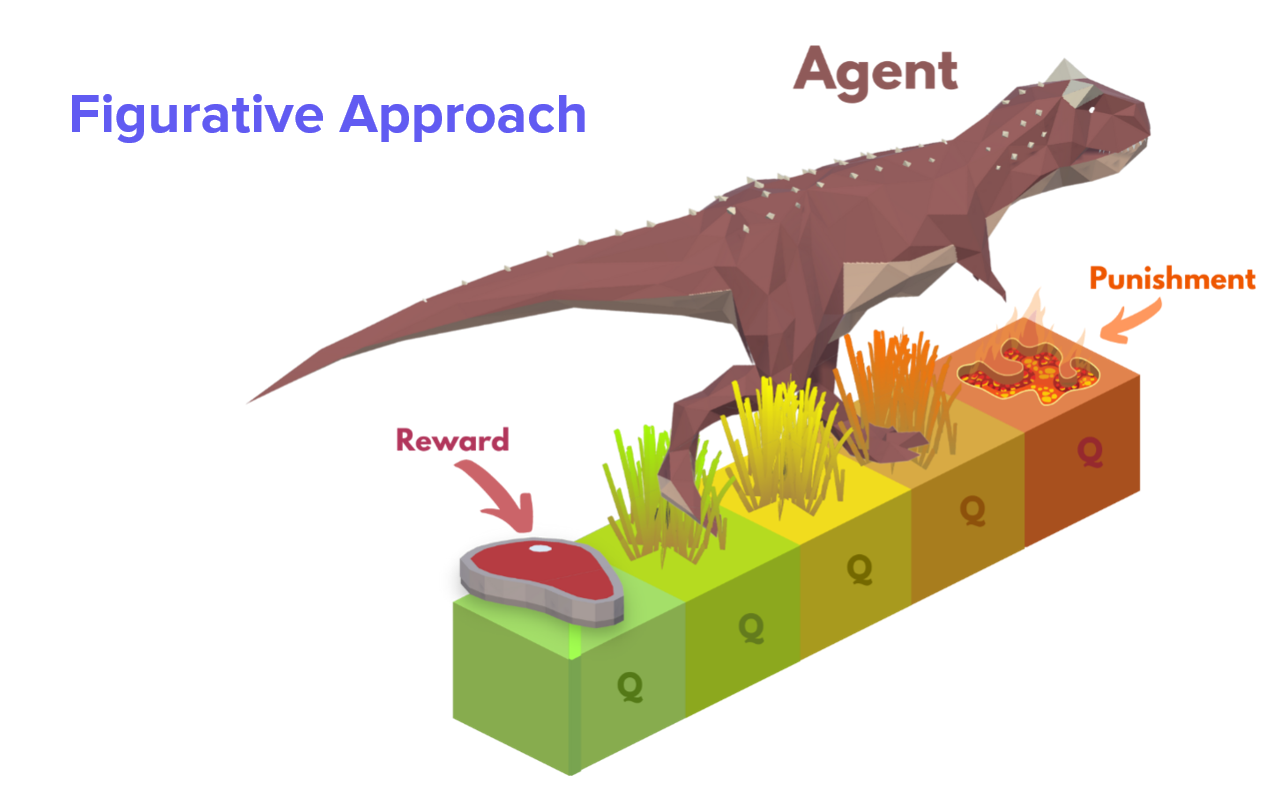}
  \caption{Figurative approach}
  \Description{a cartoon Dinosaur, meat, lava and grass}
\end{figure}
In our quest to make reinforcement learning accessible and engaging, we've meticulously crafted a series of levels that serve as windows into the diverse concepts underpinning this dynamic field. Each level acts as a microcosm, offering users a hands-on experience with distinct facets of reinforcement learning. Moreover, we've woven a simple yet captivating narrative thread throughout, leveraging the power of storytelling to imbue the learning journey with relatability and intrigue.

At the heart of our approach lies a specific hypothesis: by enveloping users in a narrative tapestry, we can cultivate a sense of connection and curiosity that primes them for deeper engagement with the world of data and machine learning. Through relatable characters, compelling plotlines, and immersive scenarios, we aim to ignite a spark of interest that propels users towards a more profound understanding of the underlying concepts.

With five meticulously designed levels, each meticulously calibrated to spotlight a different perspective of reinforcement learning, we offer users a diverse array of experiences to explore. From navigating the intricacies of reward structures to grappling with the complexities of state-action dynamics, each level presents a unique challenge that invites users to delve deeper into the nuances of the field.

Furthermore, our inclusion of a sandbox mode serves as a canvas for experimentation, allowing users to apply their newfound knowledge in a flexible and open-ended environment. Here, creativity reigns supreme as users are encouraged to devise their own strategies, test hypotheses, and witness firsthand the impact of their decisions on the learning process.

we carried out a playtest involving 20 participants. Our primary objective is to assess the participants' comprehension of the fundamental concept of Reinforcement Learning during this session. Additionally, we aim to evaluate the effectiveness of storytelling in enhancing the participants' understanding of the algorithm.

Here are our findings: Initially, all playtesters successfully associated dinosaurs with AI, meat with the goal, and lava with punishment. However, approximately 30\% of participants expressed some confusion regarding the representation of the rock. Nevertheless, all participants agreed that utilizing visual metaphors such as dinosaurs, meat, lava, and rock significantly aided their understanding of reinforcement learning. Additionally, 75\% of playtesters believed that integrating storytelling enhanced their comprehension of the algorithm.

Feedback highlighted the sandbox as the most favored feature, with players relishing the opportunity to construct their own experiences. 

In essence, our approach represents a harmonious fusion of storytelling and experiential learning, where narrative serves as a catalyst for exploration and discovery. By providing users with a rich tapestry of experiences to immerse themselves in, we aim to foster a deep-seated curiosity and enthusiasm for the world of data and machine learning, propelling them towards a journey of continuous discovery and growth.
\begin{figure}[ht]
  \centering
  \includegraphics[width=\linewidth,height=5cm]{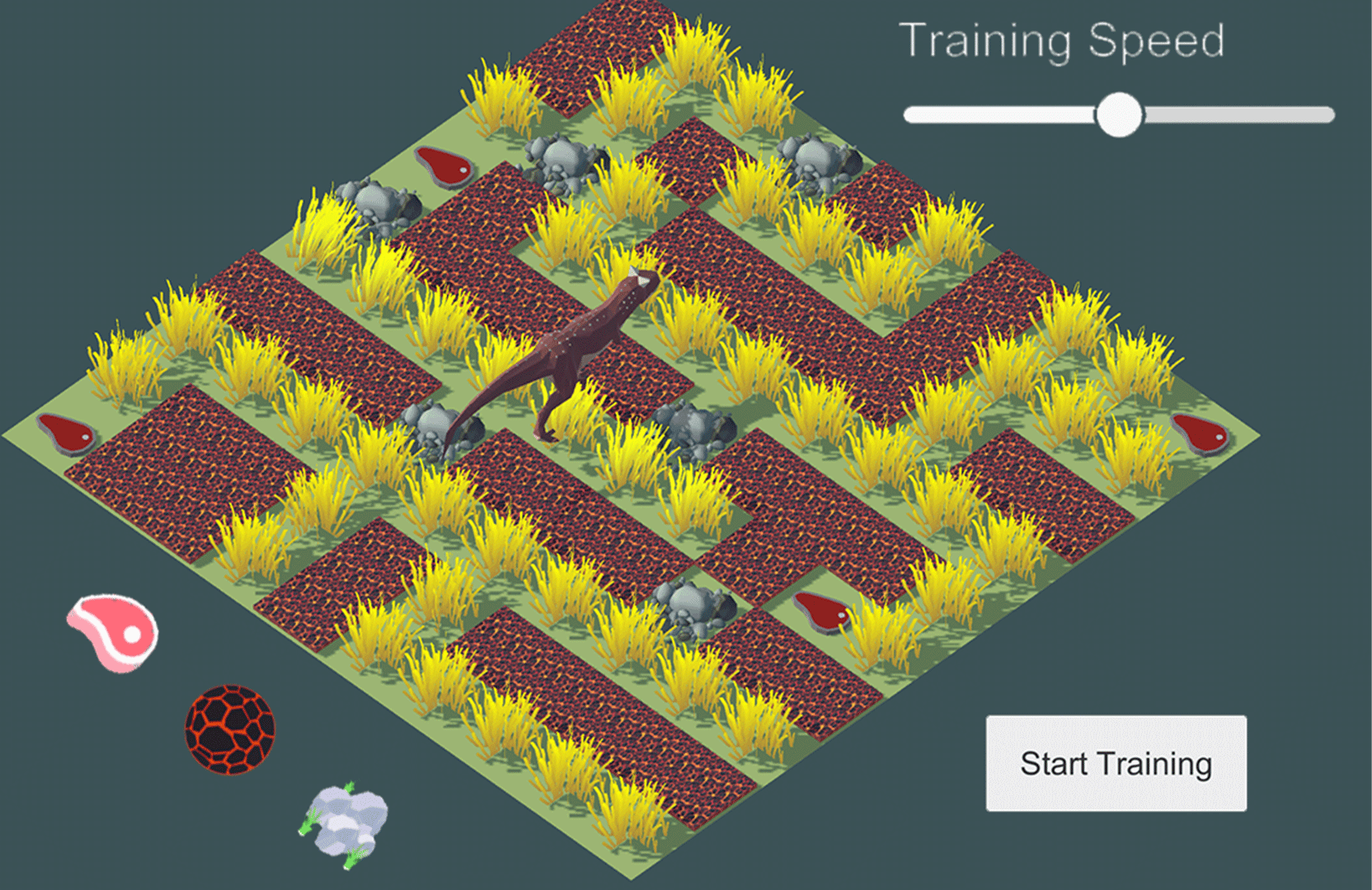}
  \caption{Reinforcement learning visualized using a dinosaur (agent), meat (reward), grass color (q-value) and lava (punishment) to represent narrative using a game format }
  \Description{Reinforcement Learning Dinosaur}
\end{figure}

The primary engagement learning centered around storytelling, which, when used with related metaphors, increased investment and inclinations towards sustained, longer-term interactions. Moreover, gamifying experience by rewarding and punishing agent behavior boosts user attentiveness\cite{Jesse19} and results in a higher content-retention rate. 

Lastly, since confusion detracts from immersion, and scaffolding prevents confusion, scaffolding became a key component of sustained engagement. Supporting user experience by providing levels, with each level focusing on a subsequent function core to reinforcement learning, fosters a better understanding of the possibilities and constraints present within a given machine learning model.

\balance

\section{Conclusion and FUTURE WORK}

Relatable, visual metaphors help increase informational and emotional accessibility. With greater accessibility comes greater engagement, creating a recursive response, feeding into a cycle resulting in more lasting paradigm shifts. 

Internally, crafting these experiences solidified our understanding of these previously abstract ML concepts, far beyond the capabilities of merely reading or being taught them in a traditional fashion. This may suggest that complex ML concepts should be taught with a tinkering mindset. If so, maker and project-based techniques may allow students to express themselves creatively using ML tools, which increase retention through experiential exploration.\cite{Mirko20}  Overall, our research would benefit from formal human studies. Additionally, the value of these tools in didactic contexts can be further explored. Such study can determine if ML experiences can provide valuable educational tools for younger people, or if tinkering with ML can broadly empower students to further engage in computer science. The haptic music prototype could also be further evaluated for potential in accessible live music transcription for the hearing-impaired.



\bibliographystyle{ACM-Reference-Format}
\bibliography{ref}

\end{document}